# Long-range four-body interactions in the Hamiltonian mean field model

Qiang Zhang [a]   Yu Xue [a†]   Haojie Luo [a]   Bingling Cen [a]

[a] Institute of Physical Science and Technology, Guangxi University, Nanning 530004, China
[†] Corresponding author: Institute of Physical Science and Technology, Guangxi University, Nanning, China.
E-mail address:

**Abstract:** In this paper, a Hamiltonian mean field model with long-range four-body interactions is proposed. The model describes a long-range mean-field system in which N unit-mass particles move on a unit circle. Each particle $\theta_i$ interacts with any three other particles through an infinite-range cosine potential with an attractive interaction ($\varepsilon>0$). By applying a method that remaps the average phase of global particle pairs onto a new unit circle, and using the saddle-point technique, the partition function is solved analytically after introducing four-body interactions, yielding expressions for the free energy f and the energy per particle U. These results were further validated through numerical simulations. The results show that the system undergoes a second-order phase transition at the critical energy $U_c$. Specifically, the critical energy corresponds to $U_c=0.32$ when the coupling constant $\varepsilon=5$, and $U_c=0.63$ when $\varepsilon=10$. Finally, we calculated the system's largest Lyapunov exponent $\lambda$ and kinetic energy fluctuations $\Sigma$ through numerical simulations. It is found that the peak of the largest Lyapunov exponent $\lambda$ occurs slightly below the critical energy $U_c$, which is consistent with the point of maximum kinetic energy fluctuations $\Sigma$. And there is a scaling law of $\Sigma/N^{1/2} \propto \lambda$ between them.

## 1. Introduction

Long-range interactions are a pervasive challenge in physics. They commonly arise in gravitational systems [1], plasma systems [2-3], and more recently, in the manipulation and observation of atomic, molecular, and optical (AMO) systems [4]. These interactions, which occur within the microscopic components of such systems, present significant theoretical and computational challenges. Despite their complexity, the fundamental role of long-range interactions in various physical phenomena makes their study indispensable[5-10]. Traditionally, a many-body system is classified as having long-range interactions if its interaction potential V(r) decays with distance r following a power law (V(r)∝r^−α), where α is a small positive constant. Such systems are notoriously difficult to analyze, both analytically and numerically. To simplify the study of long-range interactions while preserving key characteristics, a widely-used approach involves considering simplified models, such as the Hamiltonian Mean

Field (HMF) model [11-16]. The HMF model represents a system of interacting particles confined to a circular geometry. It exhibits a second-order phase transition: in the ordered phase, particles cluster together, whereas in the high-temperature phase, they are uniformly distributed along the circle. By focusing solely on the long-range nature of the interaction and ignoring their decay with distance, the HMF model enables a detailed exploration of the statistical and dynamical properties intrinsic to such systems. Notably, some results derived from the HMF model can be generalized to systems where the two-body potential decays at large distances with a power smaller than the spatial dimension [17-19].

On the other hand, research on long-range interactions has predominantly focused on two-body interactions, which have been the primary and most extensively studied aspect. In contrast, many-body interactions have received relatively little attention. The existence of many-body interactions was first confirmed among nucleons in atomic nuclei [20]. More recently, their presence has been widely acknowledged in studies of complex networks and neural networks. Many-body interactions, or higher-order interactions, involve three or more particles interacting simultaneously. Using phase reduction techniques, it has been shown that higher-order many-body interactions naturally arise in general oscillatory systems [21-22]. These interactions have a significant impact on the statistical dynamics of such systems, revealing phenomena that extend beyond the scope of conventional pairwise interaction [23-24].

In this paper, we introduce four-body interactions into the classical long-range Hamiltonian Mean Field (HMF) model and investigate its dynamics and statistical mechanics. Using a method that remaps the global particle-pair phase average onto a new unit circle, we derive expressions for the new system's partition function and free energy in the canonical ensemble. Numerical simulations are conducted using a fourth-order symplectic algorithm, providing graphical representations of the relationships between the physical quantities of interest.

The arrangement of this paper is organized as follows. Section 2 introduces four-body interactions in the classical long-range HMF model. In Section 3, we carry out the theoretical analysis and derive the solution of energy per particle U. Numerical simulation will be performed by using a fourth-order symplectic algorithm in Section 4. In Section 5, we calculate the largest Lyapunov exponent $\lambda$ and kinetic energy fluctuations $\Sigma$ of the system to study the characteristics of long-range four-body interacting system. Finally, some conclusions are given in Section 6.

## 2. The long-range four-body interacting HMF model

The Hamiltonian we consider is the following:

$$H = \sum_{i=1}^{N} \frac{p_i^{\,2}}{2} + \frac{\varepsilon}{24 N^3} \sum_{i=1}^{N} \sum_{j=1}^{N} \sum_{k=1}^{N} \sum_{l=1}^{N} \Big[ 1 - \cos\big( \theta_i + \theta_j - \theta_k - \theta_l \big) \Big] \qquad (1)$$

The first term in the equation represents the kinetic energy, while the second term corresponds to the potential energy arising from the four-body interactions. Here, $\theta_i \in [-\pi, \pi]$ denotes the position (phase) of the $i$-th particle with unit mass on the unit circle, and $p_i$ is the conjugate momentum. Each particle $\theta_i$ interacts with the other three particles. This system can be thought of as particles moving on the unit circle, interacting via a cosine potential with infinite-range attraction ($\varepsilon>0$), or equivalently as a classical XY spin model with infinite-range ferromagnetic ($\varepsilon>0$) coupling. The renormalization factor $N^3$ ensures that, in the thermodynamic limit, the energy and temperature of each particle are well-defined. We then define the order parameter of the HMF model as follows:

$$\mathbf{M} = M e^{i\varphi} = \frac{1}{N}\sum_{i=1}^{N} m_i \qquad (2)$$

In this context, $m_i = (\cos\theta_i, \sin\theta_i)$ where M and $\varphi$ represent the magnitude and phase of the order parameter, respectively. In the unit circle particle motion model, the order parameter M characterizes the degree of particle clustering. In the classical XY spin model, M represents the magnetization. In the following section, we will conduct a theoretical analysis of the model within the canonical ensemble and present the results of the corresponding numerical simulations.

## 3. Theoretical analysis

First, we reformulate the Hamiltonian of the model using a method that remaps the average phase of global particle pairs onto a new unit circle. Let $\Phi_m$ denote the average phase of $\theta_i$ and $\theta_j$, which represents the midpoint between any two position coordinates $\theta_i$ and $\theta_j$ (as illustrated in Fig.1). Using this approach, the Hamiltonian Mean Field (HMF) model on the original unit circle, which features a four-body interactions potential, can be reformulated as an HMF model on a new unit circle with a two-body interactions potential. The transformed Hamiltonian is expressed as follows:

$$
\begin{aligned}
H_S &= \sum_{i=1}^{N} N\frac{p_i^{\,2}}{2} + \frac{\varepsilon}{24N^3}\sum_{i=1}^{N}\sum_{j=1}^{N}\sum_{k=1}^{N}\sum_{l=1}^{N}\left\{1-\cos\left[\left(\theta_i+\theta_j\right)-\left(\theta_k+\theta_l\right)\right]\right\} \\
&= \sum_{m=1}^{S}\frac{p_m^{\,2}}{2} + \frac{\varepsilon}{24S}\sum_{m=1}^{S}\sum_{n=1}^{S}\left[1-\cos\left(2\Phi_m-2\Phi_n\right)\right]
\end{aligned} \qquad (3)
$$

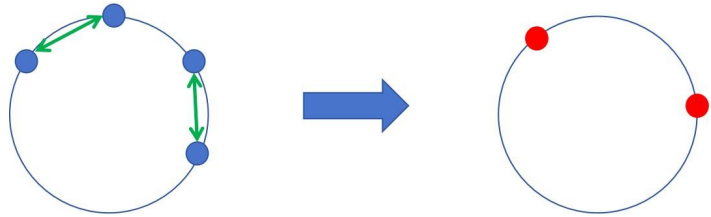

Fig.1. the schematic diagram of the average phase of global particle pairs

Before and after the transformation, the potential energy term of the system remains unchanged. Additionally, we multiply the kinetic energy term of each particle by a factor of N, ensuring that the average kinetic energy of the system remains unchanged before and after the transformation. Using this approach, we map the Hamiltonian Mean Field (HMF) model with N particles and four-body interactions on the unit circle to an HMF model with $N^2$ particles and two-body interactions on the unit circle. Next, we solve the partition function using the method described in [25]. The partition function of the system is given by:

$$Z = \int d^S p_m d^S \Phi_m \exp\left(-\beta H_S\right)$$ (4)

where $\beta=1/(k_BT)$, with $k_B$ the Boltzmann constant and T the temperature. Here, we set $k_B=1$. S is the number of particles after the transformation, where $S=N^2$. Since the average kinetic energy of the system remains unchanged before and after the transformation (3), $\beta$ also remains unchanged. Next, we define the second-order order parameter as follows:

$$\mathbf{M}_s = M_s e^{i2\phi} = \frac{1}{S}\sum_{m=1}^{S} e^{i2\Phi_m}$$ (5)

The relationship between this order parameter and the original order parameter (2) is given by:

$$\mathbf{M}_s = M_s e^{i2\phi} = \frac{1}{S}\sum_{m=1}^{S} e^{i2\Phi_m} = \frac{1}{N^2}\sum_{i,j=1}^{N} e^{i(\theta_i+\theta_j)}$$
$$= M^2 e^{i(\varphi+\varphi)} = \mathbf{M}^2$$ (6)

Substituting the Hamiltonian (3) into the partition function formula (4) and performing the integration over the momentum terms, we get:

$$Z = \left(\frac{2\pi}{\beta}\right)^{S/2} \exp\left[-\frac{\beta\varepsilon S}{24}\right] R$$ (7)

$$R = \int_{-\pi}^{\pi} d^S \Phi_m \exp\left[\frac{\beta\varepsilon}{24S}\left(\sum_{m=1}^{S} e^{i2\Phi_m}\right)^2\right]$$ (8)

By using the Gaussian integral formula [25]:

$$\exp\left(aX^2\right) = \frac{1}{\sqrt{4\pi a}}\int_{-\infty}^{+\infty} dy \exp\left(-\frac{y^2}{4a} + Xy\right)$$ (9)

We can rewrite (8) as:

$$R = \frac{6S}{\pi\beta\varepsilon}\int_{-\infty}^{+\infty}\int_{-\infty}^{+\infty} dy \exp\left[-S\left(\frac{6y^2}{\beta\varepsilon} - \ln\left(2\pi I_0\left(y\right)\right)\right)\right]$$ (10)

where $I_n$ is the modified Bessel function of order n. Finally, in the mean-field limit (i.e., as $N \rightarrow \infty$) the saddle point method can be used to calculate the integral (10). In this limit, the Helmholtz free energy f of each particle is given by:

$$\beta f = -\lim_{S\rightarrow\infty}\frac{\ln Z}{S} = -\frac{1}{2}\ln\left(\frac{2\pi}{\beta}\right) + \frac{\beta\varepsilon}{24} - \max_y\left(\frac{6y^2}{\beta\varepsilon} - \ln\left(2\pi I_0\left(y\right)\right)\right)$$ (11)

The maximum condition leads to the consistency equation:

$$\frac{I_1(y)}{I_0(y)} = \frac{12y}{\beta\varepsilon}, \mathbf{M}_s = \frac{I_1(y)}{I_0(y)} \tag{12}$$

Where $I_1(y)$ is the modified Bessel function of order 1, which is also the derivative of $I_0(y)$. The graphical solution is made easier by the fact that, for real $y>0$, the function $I_1/I_0$ is positive and monotonically increasing. We consider only the case where $\varepsilon>0$. Fig.2 presents the diagrams for two different $\beta\varepsilon$ values. For $\beta\varepsilon\leq24$ the solution is given by $y^*=0$, for $\beta\varepsilon\geq24$ the solution monotonically increases with $\beta$, and the solution $y^*=0$ is unstable. At $\beta\varepsilon=24$, a discontinuity in the second derivative of the free energy is present, indicating a second order phase transition.

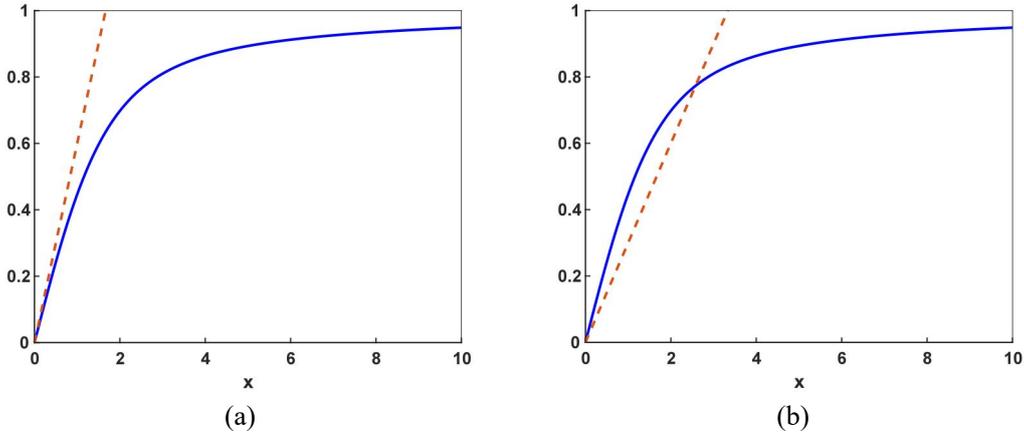

(a)                              (b)

Fig.2. The diagram of equation (12) for two $\beta\varepsilon$ values, one above the critical value and one below the critical value. (a) $\beta\varepsilon=20$, (b) $\beta\varepsilon=40$.

By taking the derivative of the free energy with respect to $\beta$, and use the relationship formula (6), we obtain the energy per particle:

$$U = \frac{\partial(\beta f)}{\partial\beta} = \frac{1}{2\beta} + \frac{\varepsilon}{24}\left(1 - \mathbf{M}^4\right) \tag{13}$$

This equation is the new analytical result obtained after introducing four-body interactions into the HMF model. In the next subsection, we will validate equation (13) through numerical simulations.

## 4. Numerical simulation

In the numerical computations, we employed a fourth-order symplectic algorithm [26-27] with a time step of 0.2, which is a high-precision numerical integration method. We performed a large number of iterations, on the order of $10^5$ steps or more, until the system reached a steady state. We analyzed the system's temperature T, order parameter M, largest Lyapunov exponent, and kinetic energy fluctuations after the system reached a steady state under various energy per particle U. The simulations were performed with particle counts N=100, 200, and 500, respectively. Fig.3 depicts the relationships between the energy per particle U and both the temperature T and the order parameter M for ε=5. According to Equation (13), a second-order phase transition occurs at β=4.8, corresponding to a critical energy $U_c$=0.32. Fig.4 presents similar relationships for ε=10. As predicted by Equation (13), a second-order phase transition takes place at β=2.4, with a critical energy of $U_c$=0.63.

The results shown in Fig.3 and Fig.4 are agreement with the predictions of Equation (13). In the following section, we will further analyze the largest Lyapunov exponent and kinetic energy fluctuations.

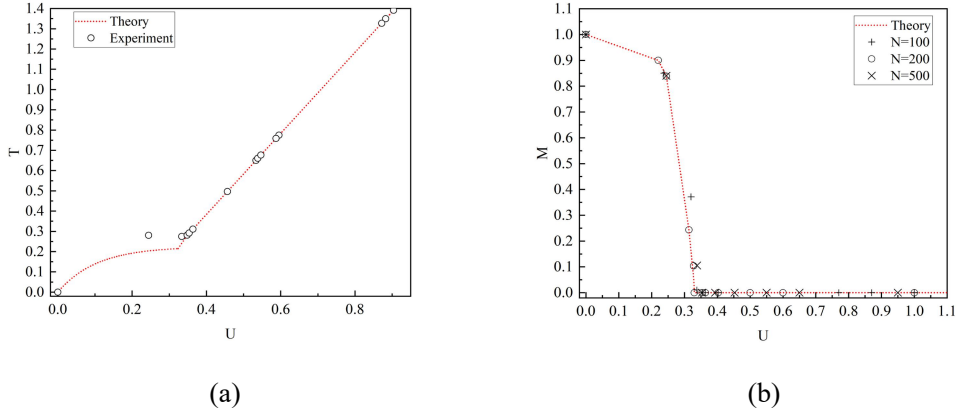

(a)                                           (b)

Fig.3. For ε=5, (a) the functional relationship between energy per particle U and temperature T. (b) the functional relationship between energy per particle U and the order parameter M. At β=4.8, corresponding to a critical energy $U_c$=0.32. The red line represents the theoretical prediction. The particle numbers are set to N=100,N=200, and N=500, respectively.

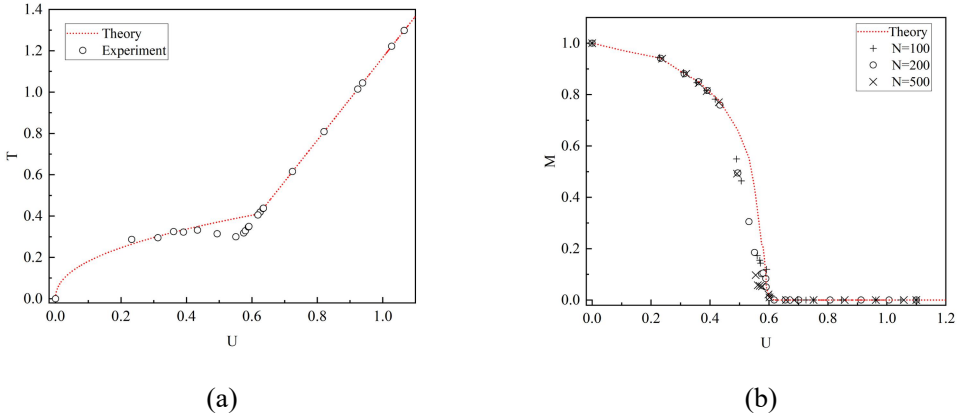

(a)                                           (b)

Fig.4. For ε=10, (a) the functional relationship between energy per particle U and temperature T. (b) the functional relationship between energy per particle U and the order parameter M. At β=2.4, corresponding to a critical energy $U_c$=0.63. The red line represents the theoretical prediction. The particle numbers are set to N=100,N=200, and N=500, respectively.

# 5. Largest Lyapunov exponent and kinetic energy fluctuations

In this section, we will discuss the chaotic characteristics of the microscopic dynamics in the four-body interactions Hamiltonian Mean Field (HMF) model. First, we numerically compute the largest Lyapunov exponent $\lambda$ using the method described in [28]. Specifically, we calculate $\lambda$ by evaluating the limit:

$$\lambda = \lim_{t \to \infty} \frac{1}{t} \ln \frac{d(t)}{d(0)} \qquad (13)$$

where:

$$d(t) = \sqrt{\sum_{i=1}^{N} \left(\delta q_i\right)^2 + \left(\delta p_i\right)^2} \qquad (14)$$

is the metric distance calculated from the infinitesimal displacement at time t. The LLE was determined using the standard method by Benettin et al. [29]. We analyzed the numerical results for system sizes N=100, N=200, and N=500. Fig.5 illustrates the functional relationship between the energy per particle U and the largest Lyapunov exponent $\lambda$ when $\varepsilon$=10. As anticipated, the largest Lyapunov exponent $\lambda$ approaches zero at both very high and very low energy per particle U, indicating quasi-integrable behavior in these regimes. However, for U>0.2, $\lambda$ exhibits a sharp, transient peak, marking a strongly chaotic region. Consistent with the findings in [28], this peak occurs at an energy U slightly below the critical energy $U_c$ and shows minimal dependence on the system size N.

On the other hand, we calculated the kinetic energy fluctuations $\Sigma$ by using the following formula:

$$\Sigma = \sqrt{\left\langle \Delta K^2 \right\rangle - \left\langle \Delta K \right\rangle^2}, \qquad \Delta K = K - \left\langle K \right\rangle \qquad (15)$$

where K represents the kinetic energy per particle. <•> indicates the time average. Fig.6 illustrates the functional relationship between energy per particle U and kinetic energy fluctuations $\Sigma$ when $\varepsilon$=10. We can observe that the behavior of $\Sigma$ is highly similar to that of $\lambda$, with both exhibiting comparable peaks in the same energy region. Additionally, the results in Fig. 7 indicate that $\Sigma/N^{1/2} \propto \lambda$. This conclusion is also supported by [30].

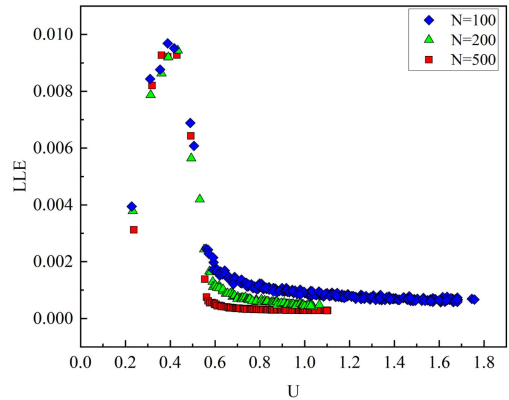

Fig.5. For $\varepsilon$=10, the functional relationship between energy per particle U and largest Lyapunov exponent $\lambda$. The particle numbers are set to N=100,N=200, and N=500, respectively.

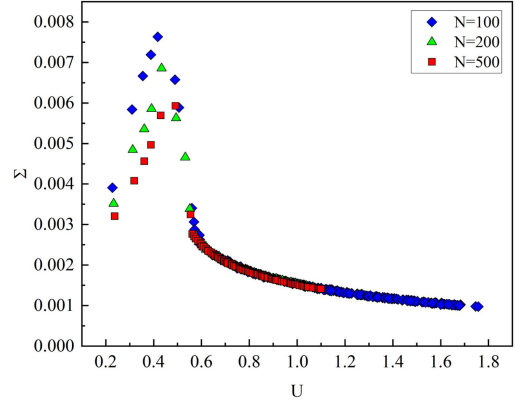

Fig.6. For ε=10, the functional relationship between energy per particle U and kinetic energy fluctuations Σ. The particle numbers are set to N=100,N=200, and N=500, respectively.

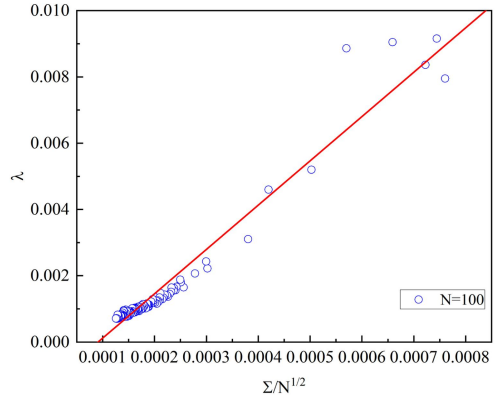

Fig.7. For ε=10 and N=100, the functional relationship between largest Lyapunov exponent λ and $\Sigma/N^{1/2}$.

# 6. Conclusions

In summary, this paper investigates a Hamiltonian mean field model with four-body interactions.This model explores a long-range mean-field system in which N unit-mass particles move on a unit circle. Each particle $\theta_i$ interacts with any three other particles through an infinite-range cosine potential with an attractive interactions (ε>0). By employing a method that remaps the average phase of global particle pairs onto a new unit circle, the original four-body interactions Hamiltonian mean-field model is reformulated as a two-body interactions Hamiltonian mean-field model on the new unit circle. Using the saddle-point method, after introducing four-body interactions, the new partition function is solved analytically, yielding expressions for the free energy f and the energy per particle U. A graphical solution reveals that a second-order phase transition occurs at βε=24. Specifically, the critical energy is $U_c$=0.32 when ε=5, and $U_c$=0.63 when ε=10.

Alternatively, we utilized a fourth-order symplectic algorithm to numerically simulate the system. The simulations were performed with a time step of 0.2, running for over $10^5$ iterations to ensure the system reached a steady state. System sizes were set to N=100, 200, and 500. We analyzed the functional relationships between the energy per particle U, temperature T, and order parameter M. The results show that for ε=5, the second-order phase transition occurs at $U_c$=0.32, while for ε=10, the transition occurs at $U_c$=0.63. These observations align well with the theoretical predictions.

Finally, we analyzed the microscopic dynamics of the model and performed numerical simulations to calculate the system's largest Lyapunov exponent and kinetic energy fluctuations Σ. We observed that the largest Lyapunov exponent λ peaks slightly below the transition energy U, which coincides with the point of maximum kinetic energy fluctuations. Away from the critical energy, both λ and Σ approach zero, with the relationship $\Sigma/N^{1/2} \propto \lambda$ holding between them. This indicates that our analytical results are supported by previous work [28]. In other words, the four-body interaction HMF model obtained through the method introduced in this paper also exhibits the same chaotic properties as the traditional HMF model.

There are still many aspects that we have not discussed here, but it can be anticipated that the methods in this paper can be extended to many-body Hamiltonian mean field models with symmetric potentials. Further research may reveal more about the dynamics and statistical mechanics behaviors of long-range interactions models with many-body interactions potentials, which will be the subject of our future research.

# Acknowledgment

The project supported by the National Natural Science Foundation of China (Grant No. 11962002) and the Innovation Project of Guangxi Graduate Education (Grant No. YCBZ2021021& YCSW2022070).